\documentclass[reprint, amssymb, amsmath, pre, showkeys]{revtex4-1}
\usepackage{graphicx, color, hyperref, nameref}
\usepackage{setspace}

\newcommand{\beq}{\begin{equation}}
\newcommand{\eeq}{\end{equation}}
\newcommand{\sat}{\mathrm{sat}}
\newcommand{\red}[1]{{#1}}

\begin{document}

\title{Tradeoffs between microbial growth phases lead to frequency-dependent and non-transitive selection}

\author{Michael Manhart}
\affiliation{Department of Chemistry and Chemical Biology, Harvard University, Cambridge, MA 02138, USA}
\author{Bharat V. Adkar}
\affiliation{Department of Chemistry and Chemical Biology, Harvard University, Cambridge, MA 02138, USA}
\author{Eugene I. Shakhnovich}
\email[To whom correspondence should be addressed.  Email: ]{shakhnovich@chemistry.harvard.edu}
\affiliation{Department of Chemistry and Chemical Biology, Harvard University, Cambridge, MA 02138, USA}
\date{\today}



\newpage{}

\begin{abstract}
Mutations in a microbial population can \red{increase} the frequency of a genotype not only by increasing its exponential growth rate, but also by decreasing its lag time or adjusting the yield (resource efficiency).  
The contribution of multiple life-history traits to selection is a critical question for evolutionary biology as we seek to predict the evolutionary fates of mutations.  Here we use a model of microbial growth to show there are two distinct components of selection corresponding to the growth and lag phases, while the yield modulates their relative importance.  The model predicts rich population dynamics when there are tradeoffs between phases: multiple strains can coexist or exhibit bistability due to frequency-dependent selection, and strains can engage in rock-paper-scissors interactions due to non-transitive selection.  We characterize the environmental conditions and patterns of traits necessary to realize these phenomena, which we show to be readily accessible to experiments.  Our results provide a theoretical framework for analyzing high-throughput measurements of microbial growth traits, especially interpreting the pleiotropy and correlations between traits across mutants.  This work also highlights the need for more comprehensive measurements of selection in simple microbial systems, where the concept of an ordinary fitness landscape breaks down.
\end{abstract}

\maketitle

\section{Introduction}

     The life history of most organisms is described by multiple traits, such as fecundity, generation time, resource efficiency, and survival probability~\cite{Orr2009}.  While all of these traits may contribute to the long-term fate of a lineage, it is often not obvious how selection optimizes all of them simultaneously, especially if there are tradeoffs~\cite{McGill2006, Warmflash2012}.  
The comparatively simple life histories of single-celled microbes make them a convenient system to study this problem.  
Microbial cells typically undergo a lag phase while adjusting to a new environment, followed by a phase of exponential growth, and finally a saturation or stationary phase when resources are depleted.  
Covariation in traits for these phases appears to be pervasive in microbial populations.  Experimental evolution of \emph{E. coli} produced wide variation of growth traits both between and within populations~\cite{Vasi1994, Novak2006}, while naturally-evolved populations of yeast showed similarly broad variation across a large number of environments~\cite{Warringer2011}.  Covariation in growth traits appears to also be important in populations adapting to antibiotics~\cite{Fitzsimmons2010, Fridman2014, RedingRoman2017, LevinReisman2017}.  Even single mutations have been found to be pleiotropic, generating variation in multiple phases~\cite{Fitzsimmons2010, Adkar2017}.

     Previous work has focused mainly on the possibility of tradeoffs between these traits, especially between exponential growth rate and yield (resource efficiency) in the context of $r$/$K$ selection~\cite{Luckinbill1978, Velicer1999, Reznick2002, Novak2006, Fitzsimmons2010, Jasmin2012a, Jasmin2012b, Bachmann2013}, as well as between growth rates at low and high concentrations of a resource~\cite{Levin1972, Stewart1973, Turner1996, Smith2011}.  However, new methods for high-throughput phenotyping of microbial populations have recently been developed to generate large data sets of growth traits~\cite{Zackrisson2016}, measuring growth rates, lag times, and yields for hundreds or thousands of strains across environmental conditions~\cite{Warringer2011}.  Some methods can even measure these traits for populations starting from single cells~\cite{LevinReisman2010, Ziv2013}.  This data requires a quantitative framework to interpret observed patterns of covariation in an evolutionary context.  For example, while growth tradeoffs have previously been proposed to cause coexistence of multiple strains~\cite{Stewart1973, Smith2011}, we lack a quantitative understanding of what patterns of traits and conditions are necessary to achieve these effects, such that they can be directly evaluated on high-throughput data.

     Here we address this problem by developing a quantitative framework for selection on multiple microbial growth traits.  We derive an expression for the selection coefficient that quantifies the relative selection pressures on lag time, growth rate, and yield.  We then determine how these selection pressures shape population dynamics over many cycles of growth, as occur in natural environments or laboratory evolution.  We find that selection is frequency-dependent, enabling coexistence and bistability of multiple strains and distorting the fixation statistics of mutants from the classical expectation.  We also find that selection can be non-transitive across multiple strains, leading to apparent rock-paper-scissors interactions.  These results are not only valuable for interpreting measurements of microbial selection and growth traits, but they also reveal how simple properties of microbial growth lead to complex population dynamics. 


\begin{figure*}
\centering\includegraphics[width=\textwidth]{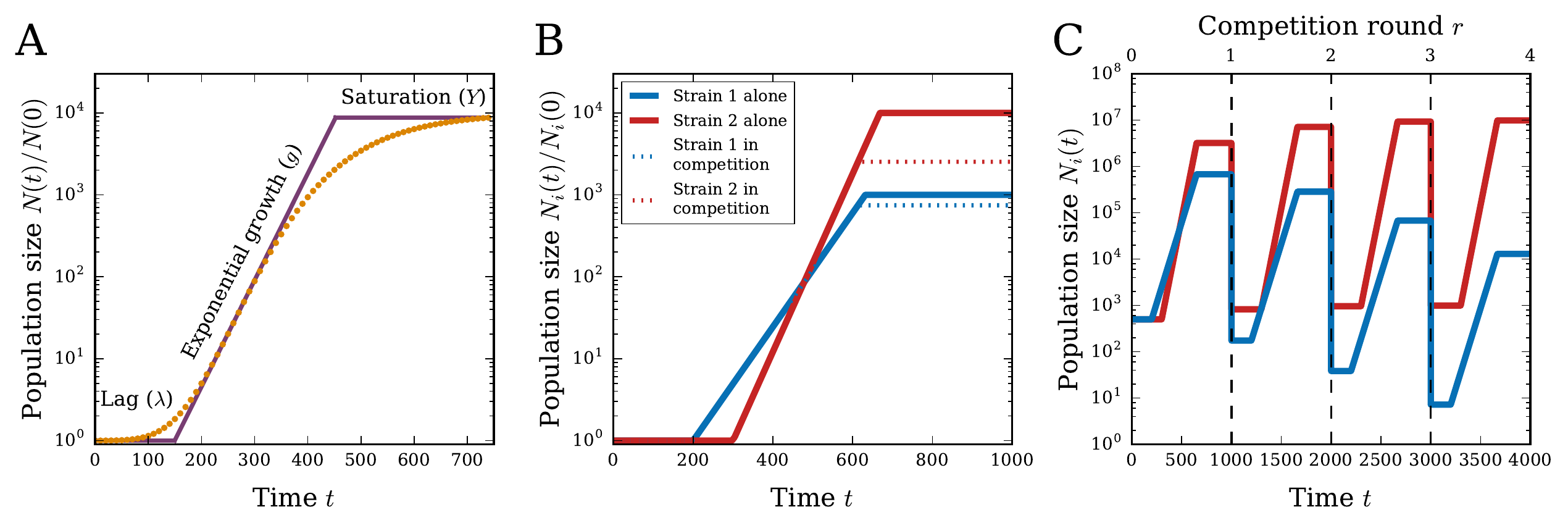}
\caption{
\textbf{Growth and selection in a microbial population.}  
(A)~Schematic of a smooth growth curve (orange points, \red{generated from a Gompertz function~\cite{Zwietering1990}}) and the minimal three-phase model (solid violet line); each phase is labeled with its corresponding growth trait.  
(B)~Two example growth curves in the three-phase model.  
Solid lines show the growth curves for each strain growing alone, while dashed lines show the growth curves of the two strains mixed together and competing for the same resources.  Note that the solid and dashed growth curves are identical until saturation, since the only effect of competition is to change the saturation time.  
(C)~Example growth curves over multiple rounds of competition.  Each vertical dashed line marks the beginning of a new growth cycle, starting from the same initial population size and amount of resources.
}
\label{fig:basic_model}
\end{figure*}

\section{Methods}

     Consider a population of microbial cells competing for a single limiting resource.  The population size $N(t)$ as a function of time (growth curve) typically follows a sigmoidal shape on a logarithmic scale, with an initial lag phase of sub-exponential growth, then a phase of exponential growth, and finally a saturation phase as the environmental resources are exhausted (Fig.~\ref{fig:basic_model}A).  We consider a minimal three-phase model of growth dynamics in which the growth curve is characterized by three quantitative traits, one corresponding to each phase of growth~\cite{Zwietering1990, Buchanan1997}: a lag time $\lambda$, an exponential growth rate $g$, and a saturation population size $N_\sat$ (Fig.~\ref{fig:basic_model}A, \red{Sec.~S1 in \emph{Supplementary Methods}}).  It is possible to generalize this model for additional phases, such as a phase for consuming a secondary resource (diauxie) or a death phase, but here we will focus on these three traits since they are most commonly reported in microbial phenotyping experiments~\cite{Warringer2011, Zackrisson2016}. 

     The saturation size $N_\sat$ depends on both the total amount of resources in the environment, as well as the cells' intrinsic efficiency of using those resources.  To separate these two components, we define $R$ to be the initial amount of the limiting resource and $Y$ to be the yield, or the number of cells per unit resource~\cite{Vasi1994}.  
Therefore $N(t)/Y$ is the amount of resources consumed by time $t$, and saturation occurs at time $t_\sat$ when $N(t_\sat) = N_\sat = RY$.  The saturation time $t_\sat$ is therefore determined intrinsically, i.e., by the growth traits of the strain, rather than being externally imposed.  It is straightforward to extend this model to multiple strains, \red{each with a distinct growth rate $g_i$, lag time $\lambda_i$, and yield $Y_i$}, and all competing for the same pool of resources (Fig.~\ref{fig:basic_model}B, \red{Sec.~S1}).  We assume different strains interact only by competing for the limiting resource; their growth traits are the same as when they grow independently.

     We focus on the case of two competing strains, such as a wild-type and a mutant.  We will denote the wild-type growth traits by $g_1, \lambda_1, Y_1$ and the mutant traits by $g_2, \lambda_2, Y_2$.  Assume the total initial population size is $N_0$ and the initial frequency of mutants is $x$.  \red{Since we are mainly interested in the relative growth of the two strains (e.g., their changes in frequency over time), only relative time scales and yields matter.}  To that end we can reduce the parameter space by using the following dimensionless quantities:

\beq
\begin{split}
\text{Relative mutant growth rate:}\quad & \gamma = (g_2 - g_1)/g_1, \\
\text{Relative mutant lag time:}\quad & \omega = (\lambda_2 - \lambda_1)g_1, \\
\text{Relative wild-type yield:}\quad & \nu_1 = RY_1/N_0, \\
\text{Relative mutant yield:}\quad & \nu_2 = RY_2/N_0. \\
\end{split}
\label{eq:param_def}
\eeq
     
\noindent Each relative yield is the fold-increase of that strain if it grows alone, starting at population size $N_0$ with $R$ resources.

     Laboratory evolution experiments, as well as seasonal natural environments, typically involve a series of these growth cycles as new resources periodically become available~\cite{Elena2003}.  We assume each round of competition begins with the same initial population size $N_0$ and amount of resources $R$, and the strains grow according to the dynamics of Fig.~\ref{fig:basic_model}B until those resources are exhausted.  The population is then diluted down to $N_0$ again with $R$ new resources, and the cycle repeats (Fig.~\ref{fig:basic_model}C).  
In each round the total selection coefficient for the mutant relative to the wild-type is

\beq
s = \log\left(\frac{N_2(t_\sat)}{N_1(t_\sat)}\right) - \log\left(\frac{N_2(0)}{N_1(0)}\right),
\label{eq:s_def}
\eeq

%

\noindent where time $t$ is measured from the beginning of the round (\red{Sec.~S2})~\cite{Crow1970, Chevin2011}.  This definition is convenient because it describes the relative change in frequency of the mutant over the wild-type during each round of competition. 
Let $x(r)$ be the mutant frequency at the beginning of the $r$th round of competition; the frequency at the end of the round will be the initial frequency $x(r+1)$ for the next round.  Using Eq.~\ref{eq:s_def}, the selection coefficient for this round is $s(x(r)) = \log(x(r+1)/[1 - x(r+1)]) - \log(x(r)/[1 - x(r)])$, which we can rearrange to obtain 

\beq
x(r+1) = \frac{x(r) e^{s(x(r))}}{1 - x(r) + x(r)e^{s(x(r))}}.
\label{eq:dynamics}
\eeq



\noindent This shows how the mutant frequency changes over rounds as a function of the selection coefficient.  If the selection coefficient is small, we can approximate these dynamics over a large number of rounds by the logistic equation: $dx/dr \approx s(x) x(1-x)$.  \red{However, for generality we use the frequency dynamics over discrete rounds defined by Eq.~\ref{eq:dynamics} throughout this work.}



\section{Results}

\subsection{Distinct components of selection on growth and lag phases}

     We can derive an approximate expression for the selection coefficient as a function of the underlying parameters in the three-phase growth model.   The selection coefficient consists of two components, one corresponding to selection on growth rate and another corresponding to selection on lag time (\red{Sec.~S3}, Fig.~S1):

\begin{subequations}
\begin{align}
s & \approx s_\text{growth} + s_\text{lag}, \label{eq:s_decomposition}\\
\intertext{where}
s_\text{growth} & = A\gamma\log\left[\frac{1}{2}H\left(\frac{\nu_1}{1-x}, \frac{\nu_2}{x}\right)\right], \nonumber\\
s_\text{lag} & = - A\omega(1 + \gamma), \label{eq:s_components}\\
A & = \frac{(1-x)/\nu_1 + x/\nu_2}{(1-x)/\nu_1 + (1 + \gamma)x/\nu_2}, \nonumber
\end{align}
\label{eq:s_formula}
\end{subequations}

\noindent and $H(a, b) = 2/(a^{-1} + b^{-1})$ denotes the harmonic mean, \red{$x$ is the frequency of the mutant at the beginning of the competition round, and $\gamma$, $\omega$, $\nu_1$, and $\nu_2$ are as defined in Eq.~\ref{eq:param_def}.}  The harmonic mean of the two yields is approximately the effective yield for the whole population (\red{Sec.~S4}).  \red{Equation~\ref{eq:s_formula} confirms that the relative traits defined in Eq.~\ref{eq:param_def} fully determine the relative growth of the strains.}

     We interpret the two terms of the selection coefficient as selection on growth and selection on lag since $s_\text{growth}$ is zero if and only if the growth rates are equal, while $s_\text{lag}$ is zero if and only if the lag times are equal.  If the mutant and wild-type growth rates only differ by a small amount ($|\gamma| \ll 1$),
then $s_\text{growth}$ is proportional to the ordinary growth rate selection coefficient $\gamma = (g_2 - g_1)/g_1$, while $-\omega = -(\lambda_2 - \lambda_1)g_1$ is the approximate selection coefficient for lag.  This contrasts with previous studies that used $\lambda~ ds/d\lambda$ as a measure of selection on lag time~\cite{Vasi1994, Wahl2015}, which assumes that selection acts on the change in lag time relative to the absolute magnitude of lag time, $(\lambda_2 - \lambda_1)/\lambda_1$.  But the absolute magnitude of lag time cannot matter since the model is invariant under translations in time, and hence our model correctly shows that selection instead acts on the change in lag time relative to the growth rate.
     

%




\begin{figure*}
\centering\includegraphics[width=\textwidth]{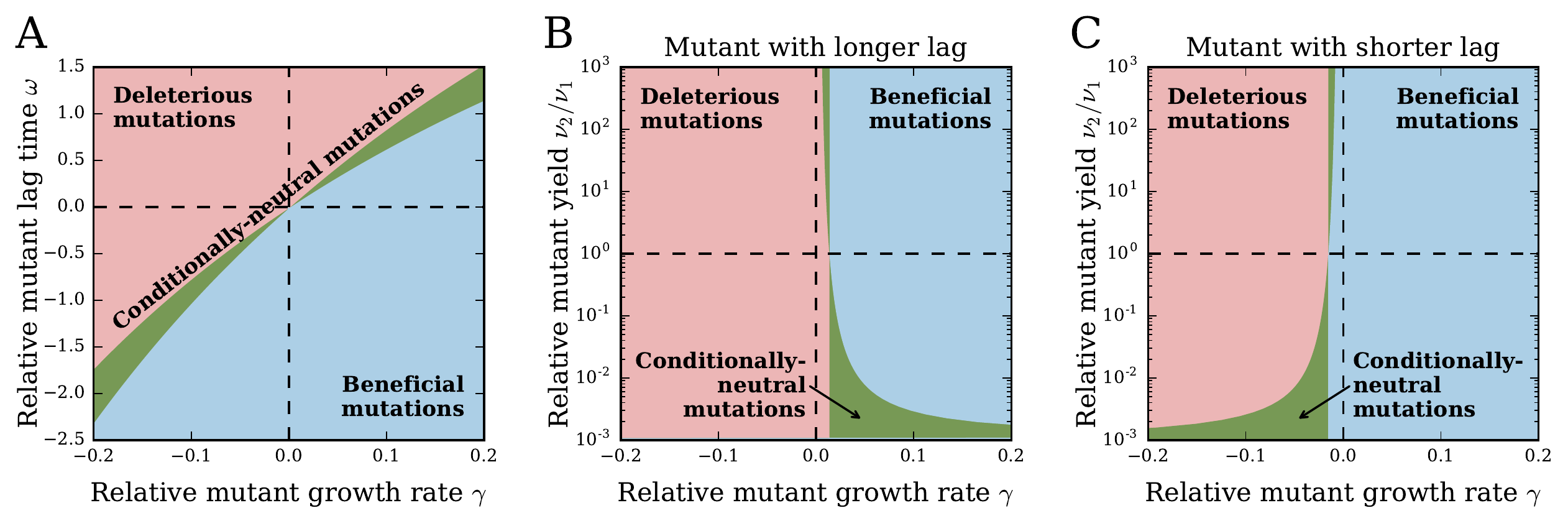}
\caption{
\textbf{Selection in mutant trait space.}   
(A)~Selection coefficient as a function of relative mutant growth rate $\gamma = (g_2 - g_1)/g_1$ and relative mutant lag time $\omega = (\lambda_2 - \lambda_1)g_1$.  
Mutants in the red region are deleterious ($s(x) < 0$) at all frequencies $x$, while mutants in the blue region are beneficial ($s(x) > 0$) at all frequencies.  Mutants in the green region are conditionally neutral, being beneficial at some frequencies and deleterious at others.  Yield values are $\nu_1 = 10^3$ and $\nu_2 = 10^4$.  
(B)~Same as (A) but in the trait space of relative mutant growth rate $\gamma$ and relative mutant yield $\nu_2/\nu_1$, for a mutant with longer lag time ($\omega = 0.1$).  
(C)~Same as (B) but for a mutant with shorter lag time ($\omega = -0.1$).
}
\label{fig:phase_diagrams}
\end{figure*}

\subsection{Effect of pleiotropy and tradeoffs on selection}

     Many mutations affect multiple growth traits simultaneously, i.e., they are pleiotropic~\cite{Fitzsimmons2010, Adkar2017}.  Given a measured or predicted pattern of pleiotropy, we can estimate its effect on selection using Eq.~\ref{eq:s_formula} (\red{Sec.~S5}).  In particular, if a mutation affects both growth and lag, then both $s_\text{growth}$ and $s_\text{lag}$ will be nonzero.  The ratio of these components indicates the relative selection on growth versus lag traits:
     
\beq
\frac{s_\text{growth}}{s_\text{lag}} = -\frac{\gamma}{\omega(1 + \gamma)}\log\left[\frac{1}{2}H\left(\frac{\nu_1}{1-x}, \frac{\nu_2}{x}\right)\right].
\label{eq:relative_selection}
\eeq

\noindent We can use this to determine, for example, how much faster a strain must grow to compensate for a longer lag time.  This also shows that we can increase the magnitude of relative selection on growth versus lag by increasing the relative yields $\nu_1$ and $\nu_2$.  Conceptually, this is because increasing the yields increases the portion of the total competition time occupied by the exponential growth phase compared to the lag phase.  Since each relative yield $\nu_i$ is proportional to the initial amount of resources per cell $R/N_0$ (Eq.~\ref{eq:param_def}), we can therefore tune the relative selection on growth versus lag in a competition by controlling $R/N_0$.  One can use this, for example, in an evolution experiment to direct selection more toward improving growth rate \red{(by choosing large $R/N_0$)} or more toward improving lag time \red{(by choosing small $R/N_0$)}.

     The ratio $s_\text{growth}/s_\text{lag}$ also indicates the type of pleiotropy on growth and lag through its sign.  If $s_\text{growth}/s_\text{lag} > 0$, then the pleiotropy is synergistic: the mutation is either beneficial to both growth and lag, or deleterious to both.  If $s_\text{growth}/s_\text{lag} < 0$, then the pleiotropy is antagonistic: the mutant is better in one trait and worse in the other.  Antagonistic pleiotropy means the mutant has a tradeoff between growth and lag.  In this case, whether the mutation is overall beneficial or deleterious depends on which trait has stronger selection.  As aforementioned, relative selection strength is controlled by the initial resources per cell $R/N_0$ through the yields (Eq.~\ref{eq:relative_selection}), so we can therefore qualitatively change the outcome of a competition with a growth-lag tradeoff by tuning $R/N_0$ to be above or below a critical value, obtained by setting $s_\text{growth} = s_\text{lag}$:
   
\beq
\text{Critical value of } \frac{R}{N_0} = \frac{2e^{\omega(1 + 1/\gamma)}}{H\left(\frac{Y_1}{1-x}, \frac{Y_2}{x}\right)}.
\label{eq:critical_RN0}
\eeq

\noindent The right side of this equation depends only on intrinsic properties of the strains (growth rates, lag times, yields) and sets the critical value for $R/N_0$, which we can control experimentally.  When $R/N_0$ is below this threshold, selection will favor the strain with the better lag time: there are relatively few resources, and so it is more important to start growing first.  On the other hand, when $R/N_0$ is above the critical value, selection will favor the strain with the better growth rate: there are relatively abundant resources, and so it is more important to grow faster.  


\subsection{Selection is frequency-dependent}

     Equation~\ref{eq:s_formula} shows that the selection coefficient $s$ depends on the initial frequency $x$ of the mutant (\red{Sec.~S6}, Fig.~S2).  This is fundamentally a consequence of having a finite resource: if resources were unlimited and selection were measured at some arbitrary time $t$ instead of $t_\sat$ (which is intrinsically determined by the strains' growth traits), then the resulting selection coefficient would not depend on $x$.

     This frequency-dependence means that some mutants are beneficial at certain initial frequencies and deleterious at others.  The traits of these ``conditionally-neutral'' mutants must satisfy
     
\beq
\min\left(\nu_1, \nu_2\right) < e^{\omega(1 + 1/\gamma)} < \max\left(\nu_1, \nu_2\right), 
\label{eq:quasineutrality}
\eeq

\noindent which is obtained by determining which trait values allow $s(\tilde{x}) = 0$ for some frequency $0 < \tilde{x} < 1$.  \red{This condition is only satisfied for mutants with a tradeoff between growth rate and lag time.}  For mutants satisfying Eq.~\ref{eq:quasineutrality}, the unique frequency at which the mutant is conditionally neutral is

\beq
\tilde{x} = \frac{\nu_1 e^{-\omega(1 + 1/\gamma)} - 1}{\nu_1/\nu_2 - 1}.
\label{eq:coexistence}
\eeq

\noindent If the mutant and wild-type have equal yields ($\nu_1 = \nu_2 = \nu$), then the mutant is neutral at all frequencies if $e^{\omega(1 + 1/\gamma)} = \nu$.  Mutants not satisfying these conditions are either beneficial at all frequencies ($s(x) > 0$) or deleterious at all frequencies ($s(x) < 0$).


\begin{figure*}
\centering\includegraphics[width=\textwidth]{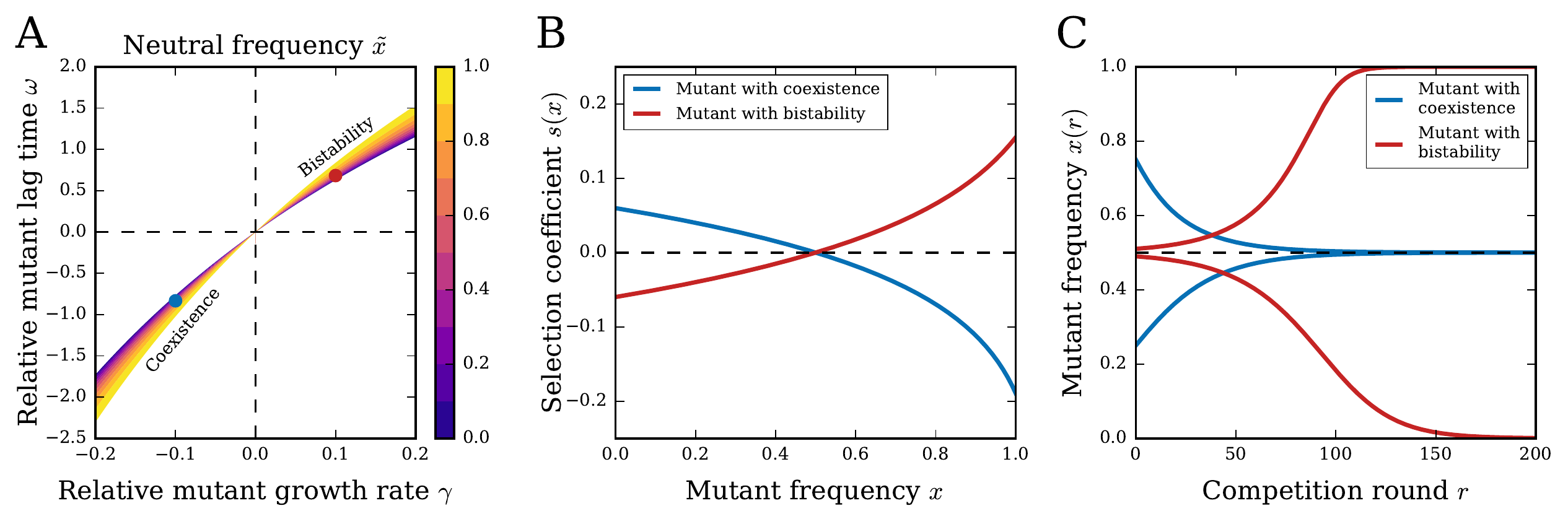}
\caption{
\textbf{Coexistence and bistability of two strains.}  
(A)~Conditionally-neutral region of growth-lag trait space where \red{coexistence or bistability occur, colored by the neutral frequency} $\tilde{x}$ (Eq.~\ref{eq:coexistence}).  Since $\nu_2 > \nu_1$ in this example, mutants in the lower branch ($\gamma < 0$) of the conditionally-neutral region coexist with the wild-type, while mutants in the upper branch ($\gamma > 0$) \red{are bistable}.  Blue and red points mark example mutants used in (B) and (C).  
(B)~Selection coefficient $s(x)$ as a function of frequency $x$ for mutants with neutral frequency $\tilde{x} = 1/2$, where one mutant has coexistence (blue) and the other \red{is bistable} (red).  
(C)~Mutant frequency $x(r)$ as a function of competition round $r$ for blue and red mutants from (A) and (B), each starting from two different initial conditions.  The black dashed line marks the neutral frequency $\tilde{x} = 1/2$.  
The yields are $\nu_1 = 10^3$ and $\nu_2 = 10^4$ in all panels. 
}
\label{fig:coexistence}
\end{figure*}

\subsection{Neutral, beneficial, and deleterious regions of mutant trait space}

     Figure~\ref{fig:phase_diagrams}A shows the regions of growth and lag trait space corresponding to conditionally-neutral (green), beneficial (blue), and deleterious (red) mutants.
The \red{slope} of the conditionally-neutral region is determined by the magnitudes of the yields: increasing both yields (e.g., by increasing the initial resources per cell $R/N_0$) makes the region \red{steeper}, since that increases relative selection on growth (Eq.~\ref{eq:relative_selection}).  

     We can further understand the role of the yields by considering the trait space of growth rate and yield (Fig.~\ref{fig:phase_diagrams}B,C), as commonly considered in $r$/$K$ selection studies~\cite{Luckinbill1978, Velicer1999, Reznick2002, Novak2006, Fitzsimmons2010, Jasmin2012a, Jasmin2012b, Bachmann2013}.  If the mutant has a longer lag time, then having a higher yield will be advantageous since the greater resource efficiency gives the mutant more time to grow exponentially to make up for its late start (Fig.~\ref{fig:phase_diagrams}B).  On the other hand, if the the mutant has a shorter lag time, then having a lower yield is better since the mutant can hoard resources 
before the wild-type grows too much (Fig.~\ref{fig:phase_diagrams}C). 
These diagrams also show there are limits to how much changes in yield can affect selection.  For example, if a deleterious mutant with slower growth ($\gamma < 0$) but shorter lag ($\omega < 0$) reduces its yield, the best it can do is become conditionally neutral (move down into the green region of Fig.~\ref{fig:phase_diagrams}C) --- it can never become completely beneficial.  Likewise, a beneficial mutant with faster growth but longer lag can never become completely deleterious by varying its yield (Fig.~\ref{fig:phase_diagrams}B).  Furthermore, a mutant with worse growth \emph{and} lag can never outcompete the wild-type, no matter how resource-efficient (high yield) it is. 
In this sense there are no pure ``$K$-strategists'' in the model~\cite{Reznick2002}.  Indeed, Eq.~\ref{eq:s_decomposition} indicates that there is no distinct selection pressure on yield, but rather it only modulates the relative selection pressures on growth versus lag.  
Note that increasing the mutant yield significantly above the wild-type value changes the selection coefficient very little, since the effective yield for the combined population (which determines the selection coefficient) is dominated by whichever strain is less efficient through the harmonic mean in Eq.~\ref{eq:s_formula}.

\subsection{Growth-lag tradeoffs enable coexistence \red{or bistability} of a mutant and wild-type}

     \red{Mutants that are conditionally neutral (satisfying Eq.~\ref{eq:quasineutrality}) due to a growth-lag tradeoff will have zero selection coefficient at an intermediate frequency $\tilde{x}$ (Eq.~\ref{eq:coexistence}).  Figure~\ref{fig:coexistence}A shows the conditionally-neutral region of trait space colored according to the neutral frequency.  For the two example mutants marked by blue and red points in Fig.~\ref{fig:coexistence}A, both with neutral frequency $\tilde{x} = 1/2$, Fig.~\ref{fig:coexistence}B shows their selection coefficients $s(x)$ as functions of frequency $x$.  Selection for the blue mutant has negative (decreasing) frequency-dependence, so that when the frequency is below the neutral frequency $\tilde{x}$, selection is positive, driving the frequency up toward $\tilde{x}$, while selection is negative above the neutral frequency, driving frequency down.  Therefore this mutant will stably coexist at frequency $\tilde{x}$ with the wild-type.  In contrast, the red mutant has positive (increasing) frequency-dependent selection, so that it has bistable long-term fates: selection will drive it to extinction or fixation depending on whether its frequency is below or above the neutral frequency.  Bistability of this type has been proposed as a useful mechanism for safely introducing new organisms into an environment without allowing them to fix unintentionally~\cite{Tanaka2017}.  Figure~\ref{fig:coexistence}C shows example trajectories of the frequencies over rounds of competitions for these two mutants.  
     Coexistence of a conditionally-neutral mutant and wild-type requires a tradeoff between growth rate and yield (Sec.~S6) --- the mutant must have faster growth rate and lower yield, or slower growth rate and higher yield --- in addition to the tradeoff between growth rate and lag time necessary for conditional neutrality.  For example, the blue mutant in Fig.~\ref{fig:coexistence} has slower growth but shorter lag and higher yield compared to the wild-type.  Therefore when the mutant is at low frequency (below $\tilde{x} = 1/2$), the overall yield of the combined population (harmonic mean in Eq.~\ref{eq:s_formula}) is approximately equal to the wild-type's yield, and since the wild-type has lower yield, this results in stronger selection on lag versus growth.  This means positive selection for the mutant, which has the shorter lag time.  In contrast, when the mutant's frequency is high, the overall yield of the population is closer to the mutant's yield, and thus there is stronger selection on growth versus lag.  This favors the wild-type strain, which has the faster growth rate, and therefore produces negative selection on the mutant.  These scenarios are reversed when the strain with faster growth (and longer lag) also has greater yield (e.g., the red mutant in Fig.~\ref{fig:coexistence}), resulting in bistability.  Since Fig.~\ref{fig:coexistence}A assumes the mutant has yield higher than that of the wild-type, all mutants in the lower branch of the conditionally-neutral region have coexistence, while all mutants in the upper branch are bistable.}
     
     
     
     Given any two strains with different yields and a tradeoff between growth and lag, in principle it is always possible to construct competition conditions such that the two strains will \red{either coexist or be bistable}.  That is, one may choose any \red{neutral} frequency $\tilde{x}$ and use Eq.~\ref{eq:critical_RN0} to determine the critical value of the initial resources per cell $R/N_0$; with $R/N_0$ set to that value, the competition will have \red{zero selection} at precisely the desired frequency.  Whether that \red{produces coexistence or bistability} depends on whether there is a tradeoff between growth and yield.  Since the bottleneck population size $N_0$ also controls the strength of stochastic fluctuations (genetic drift) between competition rounds, we can determine how to choose this parameter such that coexistence will be robust to these fluctuations (\red{Sec.~S7}).
     
     Frequency-dependent selection may also significantly distort fixation of the mutant.  In particular, it is common to measure selection on a mutant by competing the mutant against a wild-type starting from equal frequencies ($x = 1/2$)~\cite{Elena2003}.  If selection is approximately constant across all frequencies, this single selection coefficient measurement $s(1/2)$ is sufficient to accurately estimate the fixation probability and time of the mutant (\red{Sec.~S8}).  However, conditionally-neutral mutants may have fixation statistics that deviate significantly from this expectation due to frequency-dependent selection.  For example, a mutant that is neutral at $\tilde{x} = 1/2$ will have $s(1/2) = 0$ by definition, which would suggest the fixation probability of a single mutant should be the neutral value $1/N_0$.  However, its fixation probability may actually be much lower than that when accounting for the full frequency-dependence of selection (\red{Sec.~S8}, Fig.~S3).  Therefore accounting for the frequency-dependent nature of selection may be essential for predicting evolutionary fates of mutations with tradeoffs in growth traits.
     

\begin{figure*}
\centering\includegraphics[width=\textwidth]{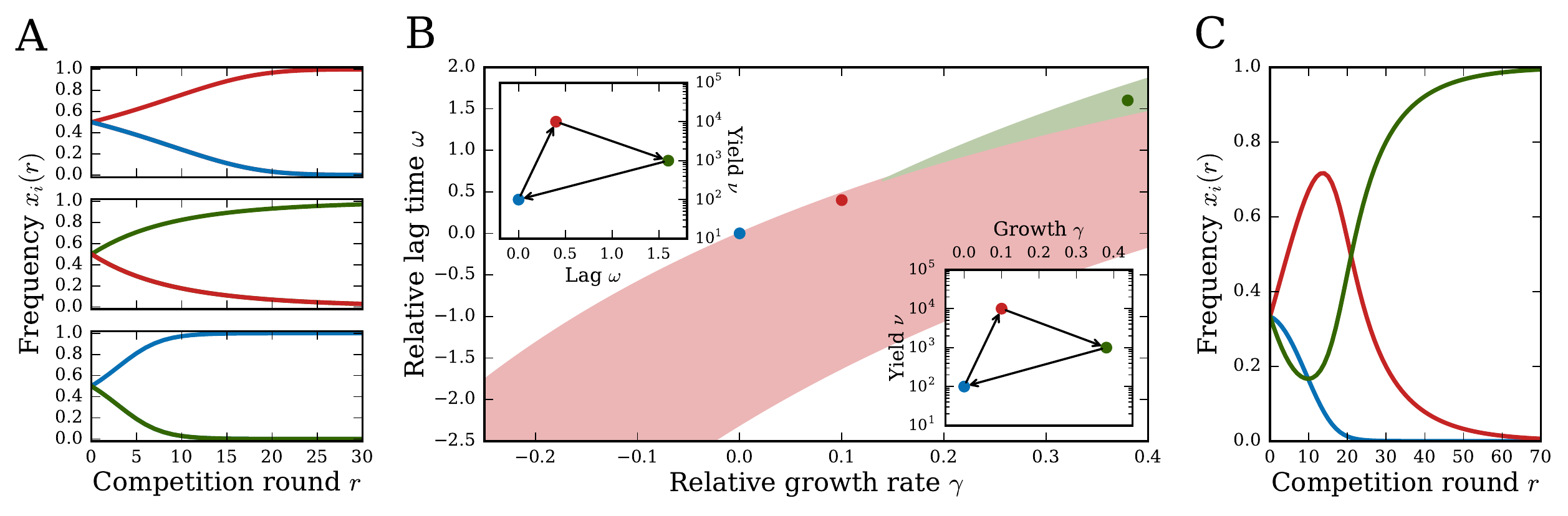}
\caption{
\textbf{Non-transitive selection over three strains.}  
(A)~An example of three strains (blue, red, green) forming a non-transitive set: in binary competitions starting from equal frequencies ($x = 1/2$), red beats blue, green beats red, and blue beats green.  
(B)~The three strains from (A) in the trait space of relative growth rate $\gamma$ and lag time $\omega$ (all relative to the blue strain); the red and green shaded regions indicate the available trait space for the red and green strains such that the three strains will form a non-transitive set.  Insets: strains in the trait space of lag time and yield $\nu$ (upper left) and trait space of growth rate and yield (lower right).  Arrows indicate which strain beats which in binary competitions.  
(C)~Dynamics of each strain's frequency $x_i(r)$ over competition rounds $r$ for all three strains in (A) simultaneously competing.  
}
\label{fig:nontransitivity}
\end{figure*}

\subsection{Selection is non-additive and non-transitive}


     We now consider a collection of many strains with distinct growth traits.  To determine all of their relative selection coefficients, in general we would need to perform binary competitions between all pairs.  However, if selection obeys the additivity condition
     
\beq
s_{ij} + s_{jk} = s_{ik},
\label{eq:additivity}
\eeq

\noindent where $s_{ij}$ is the selection coefficient of strain $i$ over strain $j$ in a binary competition, then we need only measure selection coefficients relative to a single reference strain, and from those we can predict selection for all other pairs.  
The additivity condition holds, for example, if selection coefficients are simply differences in scalar fitness values (Malthusian parameters) for each strain (i.e., $s_{ij} = f_i - f_j$).  Therefore the extent to which Eq.~\ref{eq:additivity} holds is indicative of the existence of a fitness landscape.

     Based on the selection coefficient definition (Eq.~\ref{eq:s_def}), the additivity condition would hold if \red{the selection coefficient is measured at a fixed time $t$ before saturation occurs}.  In that case, there is a scalar fitness value $f_i = g_i(t - \lambda_i)$ for each strain, and the selection coefficients are just differences in these values (\red{Sec.~S2}).  However, if we only measure selection after the finite resources are exhausted, then \red{the selection coefficient depends on the saturation time $t_\sat$, which is intrinsically determined  by the traits of the two competing strains and is therefore different for each binary competition (\red{Sec.~S4}).}  This means that the selection coefficient in this model does not obey additivity in general, although it will be approximately additive in the limit of small differences in growth traits between strains (\red{Sec.~S9}).
     
     A condition weaker than additivity is transitivity, which means that if strain 2 beats strain 1 and strain 3 beats strain 2 in binary competitions, then strain 3 must beat strain 1 in a binary competition as well~\cite{Verhoef2010}.  This must also hold for neutrality, so if strains 1 and 2 are neutral, and strains 2 and 3 are neutral, then strains 1 and 3 must also be neutral.  This essentially means that Eq.~\ref{eq:additivity} at least predicts the correct sign for each binary selection coefficient.

     If all three strains have equal yields, then selection in our model is always transitive for any initial frequencies (\red{Sec.~S10}).  If the yields are not all equal, then it is possible to find sets of three strains with non-transitive selection: each strain outcompetes one of the others in a binary competition (\red{Sec.~S10}), forming a rock-paper-scissors game~\cite{Kerr2002}.  In Fig.~\ref{fig:nontransitivity}A we show an example of three strains forming a non-transitive set.  
Figure~\ref{fig:nontransitivity}B shows the distribution of these same three strains in trait space, where the shaded regions indicate constraints on the strains necessary for them to exhibit non-transitivity.  That is, given a choice of the blue strain's traits, the red strain's traits may lie anywhere in the red shaded region, which allows the red strain to beat the blue strain while still making it possible to choose the green strain and form a non-transitive set.  Once we fix the red point, then the green strain's traits may lie anywhere in the green shaded region.

     This trait space diagram reveals what patterns of traits are conducive to generating non-transitive selection.  The trait space constraints favor a positive correlation between growth rates and lag times across strains, indicating a growth-lag tradeoff.  Indeed, these tradeoffs between growth strategies are the crucial mechanism underlying non-transitivity.  For example, in Fig.~\ref{fig:nontransitivity}A, red beats blue since red's faster growth rate and higher yield outweigh its longer lag time; green beats red due to its even faster growth rate, despite its longer lag and lower yield; and blue beats green with a shorter lag time and lower yield.  
Non-transitive strains will generally have no significant correlation between yield and growth rate or between yield and lag time (Fig.~\ref{fig:nontransitivity}B, insets); furthermore, the cycle of selective advantage through the three strains generally goes clockwise in both the lag-yield and growth-yield planes.  

     Since each strain in a non-transitive set can beat one of the others in a binary competition, it is difficult to predict \emph{a priori} the outcome of a competition with all three present.  
In Fig.~\ref{fig:nontransitivity}C we show the population dynamics for ternary competition of the non-transitive strains in Fig.~\ref{fig:nontransitivity}A,B.  
Non-transitive and frequency-dependent selection creates complex population dynamics: the red strain rises at first, while the blue and green strains drop, but once blue has sufficiently diminished, that allows green to come back (since green loses to blue, but beats red) and eventually dominate.  
Note that we do not see oscillations or coexistence in these ternary competitions, as sometime occur with non-transitive interactions~\cite{Sinervo1996, Verhoef2010}.

\section{Discussion}
\label{sec:discussion}

\subsection{Selection on multiple growth phases produces complex population dynamics}

     
     Our model shows how basic properties of microbial growth cause the standard concept of a scalar fitness landscape to break down, revealing selection to depend fundamentally on the multidimensional nature of life history. 
This occurs even for the simple periodic environment (constant $R$ and $N_0$) commonly used in laboratory evolution; fluctuating environments, as are expected in natural evolution, will likely exaggerate the importance of these effects.  
In contrast with previous theoretical work on tradeoffs between different phases of growth~\cite{Stewart1973, Smith2011}, we have obtained simple mathematical results indicating the environmental conditions and patterns of traits necessary to produce complex population dynamics such as coexistence \red{and bistability.}  In particular, we have shown how to tune the amount of resources $R$ and bottleneck population size $N_0$ such that \emph{any} pair of strains with a growth-lag tradeoff will coexist \red{or be bistable.}  
\red{In terms of ecology, this is an important demonstration of how life-history tradeoffs can enable coexistence of multiple strains even on a single limiting resource~\cite{Levin1972}.  This conflicts with the principle of competitive exclusion~\cite{Hardin1960}, which posits that the number of coexisting types cannot exceed the number of resources.  However, models that demonstrate this principle, such as the MacArthur consumer-resource model~\cite{Chesson1990}, do not account for multiple phases of life history, so that a single strain will always have overall superiority on any one resource.}

Our model furthermore provides a simple mechanism for generating non-transitive interactions, in contrast to most known mechanisms that rely on particular patterns of allelopathy~\cite{Jackson1996, Kerr2002}, morphology~\cite{Sinervo1996}, or spatial dynamics~\cite{Edwards2010}.  These results emphasize the need for more comprehensive measurements of selection beyond competition experiments against a reference strain at a single initial frequency~\cite{Elena2003}.  
As we have shown, these measurements may be insufficient to predict the long-term population dynamics at all frequencies (due to frequency-dependent selection), or the outcomes of all possible binary and higher-order competitions (due to non-transitive selection).

\subsection{Pleiotropy and correlations between traits}

     Tradeoffs among growth, lag, and yield are necessary for coexistence, \red{bistability,} and non-transitivity.  
Whether these tradeoffs are commonly realized in an evolving microbial population largely depends on the pleiotropy of mutations.  
Two theoretical considerations suggest pleiotropy between growth and lag will be predominantly synergistic. 
First, cell-to-cell variation in lag times~\cite{LevinReisman2010, Ziv2013} means that the apparent population lag time is largely governed by the cells that happen to exit lag phase first and begin dividing, which causes the population lag time to be conflated with growth rate~\cite{Baranyi1998}.   
Second, mechanistic models that attempt to explain how growth rate and lag time depend on underlying cellular processes also predict synergistic pleiotropy~\cite{Baranyi1994, Swinnen2004, Himeoka2017}; conceptually, this is because the product of growth rate and lag time should be a positive constant corresponding to the amount of metabolic ``work'' 
that the cell must perform to exit lag and begin to divide.  Pleiotropy between growth rate and yield, on the other hand, is generally expected to be antagonistic 
due to thermodynamic constraints between the rate and yield of metabolic reactions~\cite{Pfeiffer2001, MacLean2007}, although this constraint may not necessarily induce a correlation~\cite{Wong2009}.

     Distributions of these traits have been measured for both bacteria and fungi.  Correlations between growth rate and yield have long been the focus of $r$/$K$ selection studies; some of these experiments have indeed found tradeoffs between growth rate and yield~\cite{MacLean2007, Jasmin2012a, Jasmin2012b, Bachmann2013}, but others have found no tradeoff, or even a positive correlation~\cite{Luckinbill1978, Velicer1999, Novak2006, Fitzsimmons2010, Warringer2011}.  Measurements of lag times have also found mixed results~\cite{Swinnen2004, Warringer2011, Himeoka2017, Wang2015, Adkar2017}.  
However, most of these data are for evolved populations, which may not reflect the true pleiotropy of mutations: distributions of fixed mutations may be correlated by selection even if the underlying distributions of mutations are uncorrelated.  Our model shows that higher yield is only beneficial for faster growth rates, and so selection will tend to especially amplify mutations that increase both traits, which may explain some of the observed positive correlations between growth rate and yield.  Indeed, data on the distributions of growth rates and yields from individual clones \emph{within} a population show a negative correlation~\cite{Novak2006}.  The model developed here will be useful for further exploring the relationship between the underlying pleiotropy of mutations and the distribution of traits in evolved populations.

\subsection{Analysis of experimental growth curves and competitions}
     
     Given a collection of microbial strains, we can measure their individual growth curves and determine growth rates, lag times, and yields.  In principle, we can use the model (Eq.~\ref{eq:s_formula}) to predict the outcome of any binary competition with these strains.  \red{These strains need not be mutants of the same species, as we primarily discuss here, but can even be different species.}  In practice, however, there are several challenges in applying the model to this data.  First, real growth dynamics are undoubtedly more complicated than the minimal model used here.  There are additional time scales, such as the rate at which growth decelerates as resources are exhausted~\cite{Stewart1973}; other frequency-dependent effects, such as a dependence of the lag time on the initial population size~\cite{Kaprelyants1996}; and more complex interactions between cells, such as cross-feeding~\cite{Turner1996}, especially between different species.  In addition, the measured traits and competition parameters may be noisy, due to intrinsic noise within the cells as well as the extrinsic noise of the experiment.  
     
     Nevertheless, the simplicity of the model investigated here makes it a useful tool for identifying candidate strains from a collection of individual growth curves that may have interesting dynamics in pairs or in multi-strain competitions, which can then be subsequently tested by experiment.  Existing technologies enable high-throughput measurement of individual growth curves for large numbers of strains~\cite{LevinReisman2010, Ziv2013, Zackrisson2016}, but systematic measurements of competitions are limited by the large number of possible strain combinations, as well as the need for sequencing or fluorescent markers to distinguish strains.  The model can therefore help to target which competition experiments are likely to be most interesting by computationally scanning all combinations and setting bounds on various parameters to be compared with experimental uncertainties.  For example, we can identify pairs of strains with growth-lag tradeoffs and predict a range of competition conditions $R/N_0$ that will lead to coexistence.  
We can also identify candidate sets of strains for demonstrating non-transitive selection.  
\red{Even for sets of strains with additional interactions beyond competition for a single resource, which will almost certainly be the case when the strains are different species, our results can serve as a null model for testing the importance of these other interactions, beyond variation in growth traits, in shaping population dynamics.}


\begin{acknowledgments}
     We thank Tommaso Biancalani, Parris Humphrey, and William Jacobs for valuable discussions, and Tal Einav for a critical reading of the manuscript.  This work was supported by NIH awards F32 GM116217 to MM and R01 GM068670 to EIS.
\end{acknowledgments}
     
\section*{Author contributions}

     MM, BVA, and EIS designed research; MM and BVA carried out calculations and analyzed data; MM wrote the manuscript.  All authors edited and approved the final version.


\bibliographystyle{vancouver}
\bibliography{References}


\end{document}


\title{Supplementary Methods: \\ Tradeoffs between microbial growth phases lead to frequency-dependent and non-transitive selection}

\author{Michael Manhart}
\affiliation{Department of Chemistry and Chemical Biology, Harvard University, Cambridge, MA 02138, USA}
\author{Bharat V. Adkar}
\affiliation{Department of Chemistry and Chemical Biology, Harvard University, Cambridge, MA 02138, USA}
\author{Eugene I. Shakhnovich}
\affiliation{Department of Chemistry and Chemical Biology, Harvard University, Cambridge, MA 02138, USA}



\newpage{}

\maketitle

\setcounter{page}{1}
\setcounter{equation}{0}
\setcounter{figure}{0}
\setcounter{section}{0}
\setcounter{subsection}{0}
\renewcommand{\theequation}{S\arabic{equation}}
\renewcommand{\thefigure}{S\arabic{figure}}
\renewcommand{\thesection}{S\arabic{section}}

\section{Minimal three-phase model of population growth}  
     
     Let each strain $i$ have lag time $\lambda_i$, growth rate $g_i$, and initial population size $N_i(0)$, so that its growth dynamics obey (Fig.~1A)

\beq
N_i(t) = \left\{
\begin{array}{ll}
N_i(0) & 0 \leq t < \lambda_i, \\
N_i(0) e^{g_i(t - \lambda_i)} & \lambda_i \leq t < t_\sat, \\
N_i(0) e^{g_i(t_\sat - \lambda_i)} & t \geq t_\sat. \\
\end{array}
\right.
\label{eq:growth_model}
\eeq

\noindent The time $t_\sat$ at which growth saturates is determined by a model of resource consumption.  Let $R$ be the initial amount of resources.  We assume that each strain consumes resources in proportion to its population size, for example, if the limiting resource is space.  Let the yield $Y_i$ be the number of cells of strain $i$ per unit of the resource.  Therefore the resources are exhausted at time $t = t_\sat$ such that

\beq
\sum_i \frac{N_i(t_\sat)}{Y_i} = R.
\label{eq:saturation_condition}
\eeq

 


     We can alternatively assume that each strain consumes resources in proportion to its total number of cell divisions, rather than its total number of cells.  The number of cell divisions for strain $i$ that have occurred by time $t$ is $N_i(t) - N_i(0)$.  Redefining the yield $Y_i$ as the number of cell divisions per unit resource, saturation now occurs at the time $t = t_\sat$ satisfying 
     
\beq
\sum_i \frac{N_i(t_\sat) - N_i(0)}{Y_i} = R.
\eeq

\noindent For simplicity we use the first model (Eq.~\ref{eq:saturation_condition}) throughout this work, but it is straightforward to translate all results to the second model using the transformation $R \to R + \sum_i N_i(0)/Y_i$.  This correction will generally be small, though, since $\sum_i N_i(0)/Y_i$ is the amount of resources that the initial population of cells consume for their first divisions, and this amount will usually be much less than the total resources $R$.  It is also straightforward to further generalize this model to include other modes of resource consumption, such as consuming the resource per unit time during lag phase.


\section{Definition of selection coefficient}

     The selection coefficient per unit time is
    
\beq
\sigma(t) = \frac{d}{dt} \log\left(\frac{N_2(t)}{N_1(t)}\right).
\label{eq:instant_s_def}
\eeq

\noindent In the minimal three-phase growth model (Eq.~\ref{eq:growth_model}), we can write the growth curve as $N_i(t) = N_i(0) e^{g_i(t - \lambda_i)\Theta(t - \lambda_i)}$, where $\Theta(t)$ is the Heaviside step function.  Then the instantaneous selection coefficient is:

\beq
\begin{split}
\sigma(t) & = \frac{d}{dt} \bigg[g_2(t - \lambda_2)\Theta(t - \lambda_2) - g_1(t - \lambda_1)\Theta(t - \lambda_1) \bigg] \\
& = g_2\Theta(t - \lambda_2) - g_1\Theta(t - \lambda_1), \\
\end{split}
\eeq

\noindent for $t < t_\sat$, and $\sigma(t) = 0$ for $t > t_\sat$.

     Since we are mainly concerned with how the mutant frequency changes over whole cycles of growth, it is more convenient to integrate this instantaneous selection coefficient to obtain the total selection coefficient per cycle: \red{

\beq
\begin{split}
s & = \int_0^{t_\sat} dt~ \sigma(t) \\
& = g_2(t_\sat - \lambda_2)\Theta(t_\sat - \lambda_2) \\
& \qquad\qquad\qquad - g_1(t_\sat - \lambda_1)\Theta(t_\sat - \lambda_1),
\end{split}
\label{eq:s_def_alternate1}
\eeq

\noindent which, using Eq.~\ref{eq:instant_s_def}, is equivalent to the definition in \eqsdef~from the main text.  If we exclude the trivial case where the time to saturation is less than one of the lag times (so that one strain does not grow at all), the selection coefficient simplifies to
     
\beq
s = g_2(t_\sat - \lambda_2) - g_1(t_\sat - \lambda_1).
\label{eq:s_def_alternate2}
\eeq

}



\section{Derivation of selection coefficient expression}

     To determine how $s$ explicitly depends on the underlying parameters, we must solve the saturation condition in Eq.~\ref{eq:saturation_condition} for $t_\sat$:

\beq
R = \frac{N_0x_1}{Y_1}e^{g_1(t_\sat - \lambda_1)} + \frac{N_0x_2}{Y_2}e^{g_2(t_\sat - \lambda_2)},
\label{eq:pairwise_saturation}
\eeq

\noindent where $N_0$ is the total initial population size and $x_1, x_2$ are the initial frequencies of the wild-type and mutant.  We ignore the trivial case where one strain saturates before the other starts to grow.  While we cannot analytically solve this equation in general, we can obtain a good approximation in the limit of weak selection ($|s| \ll 1$).  We first rewrite the equation in terms of the selection coefficient using Eq.~\ref{eq:s_def_alternate2}:

\beq
R = N_0 e^{g_1(t_\sat - \lambda_1)} \left(\frac{x_1}{Y_1} + \frac{x_2}{Y_2} e^s \right).
\eeq

\noindent We then solve for $t_\sat$ and expand to first order in $s$:

\begin{multline}
t_\sat \approx \lambda_1 - \frac{1}{g_1} \log\left[\frac{N_0}{R}\left(\frac{x_1}{Y_1} + \frac{x_2}{Y_2}\right)\right] \\ 
- \frac{x_2/Y_2}{g_1(x_1/Y_1 + x_2/Y_2)}s.
\end{multline}

\noindent For self-consistency this expression for $t_\sat$ should be invariant under exchange of the mutant and wild-type indices and switching the sign of $s$; equating these two equivalent expressions for $t_\sat$ allows us to solve for $s$, which gives the main result in \eqsformula.

\begin{figure}
\centering\includegraphics[width=0.9\columnwidth]{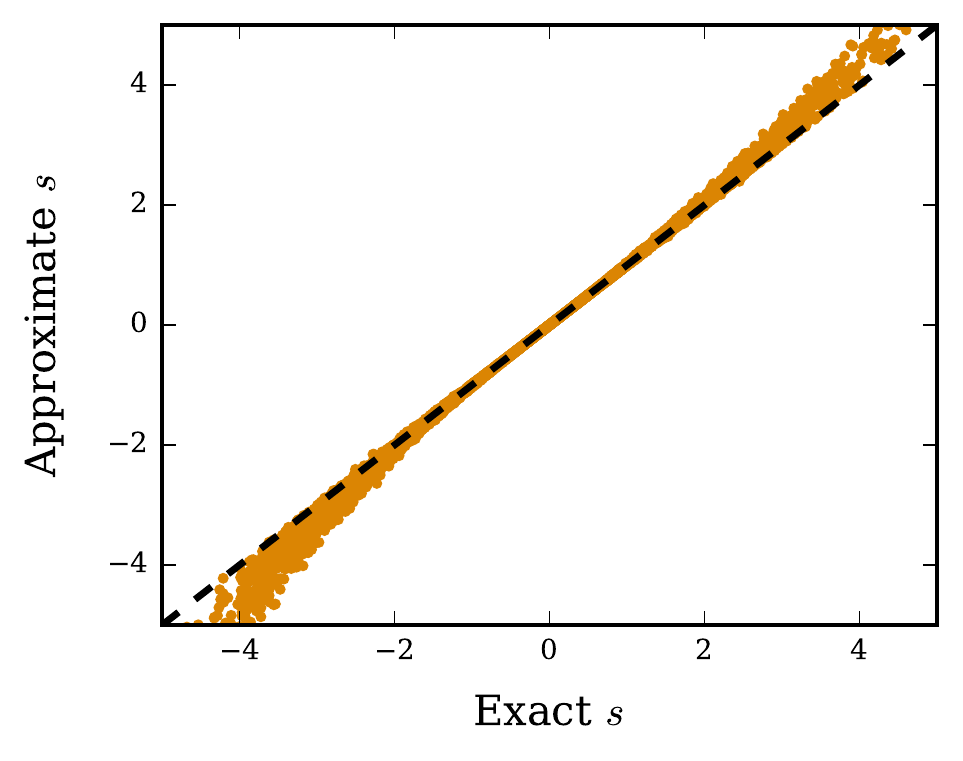}
\caption{
\textbf{Test of selection coefficient approximation.}  
Comparison of the approximate selection coefficient formula (\eqsformula) with the exact result obtained using the definition in Eq.~\ref{eq:s_def_alternate1} and a numerical solution to the saturation equation (Eq.~\ref{eq:pairwise_saturation}).  Each orange point corresponds to a different set of randomly-generated growth traits ($\gamma$, $\omega$, $\nu_1$, $\nu_2$; see \eqparamdef) and initial mutant frequencies $x$.  The black dashed line is the line of identity.
}
\label{fig:approximation_test}
\end{figure}

     In Fig.~\ref{fig:approximation_test} we compare the selection coefficient calculated from this approximate expression with the exact result obtained by numerically solving Eq.~\ref{eq:pairwise_saturation} for $t_\sat$ and then directly calculating $s$ using the definition of Eq.~\ref{eq:s_def_alternate1}.  This empirically shows that although the derivation relies on the approximation of weak selection ($|s| \ll 1$), \eqsformula~is extremely accurate over a wide range of parameter values, even up to rather strong selection strengths $|s| \sim 1$.    Furthermore, the expression is exact in two special cases: when the mutant and the wild-type are selectively neutral ($s = 0$), and when the mutant and wild-type have equal growth rates ($g_1 = g_2 = g$), since $s = -(\lambda_2 - \lambda_1)g = -\omega$ according to Eq.~\ref{eq:s_def_alternate2}.






\section{Saturation time and total population size}

     Here we derive expressions for the saturation time $t_\sat$ and the total population size at saturation

\beq
\begin{split}
N_\sat & = N_1(t_\sat) + N_2(t_\sat) \\
& = N_0x_1 e^{g_1(t_\sat - \lambda_1)} + N_0x_2 e^{g_2(t_\sat - \lambda_2)}. \\
\end{split}
\label{eq:Nsat}
\eeq
     
\noindent We again assume the nontrivial case of $t_\sat > \lambda_1, \lambda_2$.  First, if the growth rates are equal ($g_1 = g_2 = g$), we can obtain exact solutions since the two-strain saturation condition (Eq.~\ref{eq:pairwise_saturation}) is analytically solvable for $t_\sat$:
     
\beq
\begin{split}
t_\sat & = \frac{1}{g} \log\left[\frac{R}{2N_0}H\left(\frac{Y_1e^{g\lambda_1}}{x_1}, \frac{Y_2e^{g\lambda_2}}{x_2}\right)\right], \\
N_\sat & = \frac{1}{2} (x_1 e^{-g\lambda_1} + x_2 e^{-g\lambda_2}) H\left(\frac{RY_1e^{g\lambda_1}}{x_1}, \frac{RY_2e^{g\lambda_2}}{x_2}\right). \\
\end{split}
\eeq

%

     If the growth rates are unequal ($g_1 \neq g_2$), then we must rely on the small $s$ approximation.  We can rearrange Eq.~\ref{eq:s_def_alternate2} to obtain $t_\sat$ as a function of $s$:
     
\beq
t_\sat = \frac{s + g_2\lambda_2 - g_1\lambda_1}{g_2 - g_1}.
\label{eq:s_vs_tsat}
\eeq
     
\noindent We can then substitute the approximate expression for $s$ (\eqsformula) into Eq.~\ref{eq:s_vs_tsat}:  

\begin{multline}
t_\sat \approx \left(\frac{x_1/Y_1 + x_2/Y_2}{g_1x_1/Y_1 + g_2x_2/Y_2}\right) \left(\log\left[\frac{R}{2N_0}H\left(\frac{Y_1}{x_1}, \frac{Y_2}{x_2}\right)\right] \right. \\
\left. - \frac{g_1g_2(\lambda_2 - \lambda_1)}{g_2 - g_1}\right) + \frac{g_2\lambda_2 - g_1\lambda_1}{g_2 - g_1}.
\end{multline}
 
\noindent To obtain an expression for $N_\sat$ in this approximation, we rewrite its definition (Eq.~\ref{eq:Nsat}) in terms of $s$ using Eq.~\ref{eq:s_vs_tsat}:

\begin{multline}
N_\sat = N_0 e^{g_1g_2(\lambda_2 - \lambda_1)/(g_2 - g_1)} \\
\times \left(x_1 e^{g_1s/(g_2 - g_1)} + x_2 e^{g_2s/(g_2 - g_1)}\right).
\label{eq:Nsat_vs_s}
\end{multline}

\noindent For small $s$, we can show from \eqsformula~that

\begin{widetext}
\beq
\begin{split}
e^{g_1g_2(\lambda_2 - \lambda_1)/(g_2 - g_1)} & \approx \frac{1}{2N_0} H\left(\frac{RY_1}{x_1}, \frac{RY_2}{x_2}\right) \exp\left[-\left(\frac{s}{g_2 - g_1}\right)\left(\frac{g_1x_1/Y_1 + g_2x_2/Y_2}{x_1/Y_1 + x_2/Y_2}\right)\right] \\
& \approx \frac{1}{2N_0} H\left(\frac{RY_1}{x_1}, \frac{RY_2}{x_2}\right) \left[1 - \left(\frac{s}{g_2 - g_1}\right)\left(\frac{g_1x_1/Y_1 + g_2x_2/Y_2}{x_1/Y_1 + x_2/Y_2}\right)\right]. \\
\end{split}
\eeq
\end{widetext}

\noindent Substituting this into Eq.~\ref{eq:Nsat_vs_s} and expanding to first order in $s$ shows that the saturation population size is approximately

\begin{multline}
N_\sat \approx \frac{1}{2} H\left(\frac{RY_1}{x_1}, \frac{RY_2}{x_2}\right) \\
\times \left[1 - \frac{x_1x_2(Y_2^{-1} - Y_1^{-1})}{x_1/Y_1 + x_2/Y_2}s\right].
\end{multline}


     

\noindent Therefore the saturation size in the neutral case ($s = 0$) is

\beq
\begin{split}
N_\sat & = \frac{1}{2} H\left(\frac{RY_1}{x_1}, \frac{RY_2}{x_2}\right) \\
& = H\left(\frac{N_{\sat,1}}{2x_1}, \frac{N_{\sat,2}}{2x_2}\right), \\
\end{split}
\eeq

\noindent since $RY_i = N_{\sat,i}$, where $N_{\sat,i}$ is the saturation population size of strain $i$ if no other strains are present.  So for a neutral pair of strains, the total population grows to the harmonic mean of the saturation population sizes of the individual strains; this shows that we can interpret the harmonic mean of both strains' yields as the effective yield for the combined population.  When selection is nonzero, the effective yield is perturbed above this value if the strain with higher yield is also positively selected (e.g., $Y_2 > Y_1$ and $s > 0$), while otherwise it is perturbed below the neutral value.


%
%
%

%
%
%
%
%
%



\section{Effect of correlated pleiotropy on selection}

     Mutational effects on growth traits may not only be pleiotropic, but they may also be correlated.  The simplest case is a linear correlation between growth traits across many mutations or strains:

\beq
\lambda \approx \frac{a}{g} + \text{constant}, \qquad \nu \approx bg + \text{constant},
\label{eq:correlations}
\eeq

\noindent where $a$ and $b$ are proportionality constants.  We take lag time to be linearly correlated with growth time (reciprocal growth rate), rather than growth rate, since then both traits have units of time and the constant $a$ is dimensionless.  Various models predict linear correlations of this form~\cite{Baranyi1994, Baranyi1998, Pfeiffer2001, Swinnen2004, MacLean2007, Himeoka2017}, which have been tested on measured distributions of traits~\cite{Novak2006, MacLean2007, Fitzsimmons2010, Warringer2011, Jasmin2012a, Jasmin2012b, Bachmann2013} (see Discussion in main text). 

     We can combine this model with the selection coefficient in \eqsformula~to quantify how much selection is amplified or diminished by correlated pleiotropy.  That is, if a mutation changes growth rate by a small amount $\Delta g = g_2 - g_1$ from the wild-type, then according to Eq.~\ref{eq:correlations} it will also change lag time by $\Delta\lambda \approx -a\Delta g/g^2$ and yield by $\Delta\nu = b\Delta g$, and hence the expected selection coefficient will be (using $\gamma = \Delta g/g$)

\beq
s \approx \gamma(\log \nu + a).
\label{eq:s_with_correlations}
\eeq

\noindent This shows that correlations between growth and yield have no effect on selection to leading order, since selection only depends logarithmically on yield.  Correlations between growth and lag, however, can have a significant amplifying or diminishing effect.  Since $\log\nu > 0$ always, synergistic pleiotropy ($a > 0$) will tend to increase the magnitude of selection, while antagonistic pleiotropy ($a < 0$) will tend to reduce it.  The significance of this effect depends on the relative value of $a$ compared to $\log\nu$; in general, the logarithm and the dimensionless nature of $a$ suggest both should be order 1 and therefore comparable.



\section{Frequency-dependence of selection}

     The selection coefficient in \eqsformula~depends on the initial mutant frequency $x$.  Here we show that $s(x)$ is a monotonic function of the frequency $x$; this is important because it means that conditional neutrality ($s(\tilde{x}) = 0$) occurs at a unique frequency $\tilde{x}$ (\eqcoexistence).  We use an exact argument starting from the original model because the approximate $s(x)$ function in \eqsformula~has spurious non-monotonic behavior in some regimes.  For simplicity we again use the dimensionless growth parameters defined in \eqparamdef.  

     If the mutant and wild-type have equal growth rates ($\gamma = 0$), then we have previously showed that $s(x) = -\omega$, so it is constant (and hence monotonic) in $x$.  Now we consider $\gamma \neq 0$.  In this case we can write the saturation condition in terms of $s(x)$ by substituting Eq.~\ref{eq:s_vs_tsat} for $t_\sat$ in Eq.~\ref{eq:pairwise_saturation}:

\beq
e^{\omega(1 + \gamma)/\gamma}\left[\frac{(1-x)}{\nu_1} e^{s(x)/\gamma} + \frac{x}{\nu_2} e^{(1 + 1/\gamma)s(x)}\right] = 1.
\label{eq:saturation_condition_s}
\eeq

\noindent We can differentiate with respect to $x$ and solve for $ds/dx$ to obtain the differential equation

\beq
\frac{ds}{dx} = \frac{\gamma\left(1 - e^{s(x)}\nu_1/\nu_2\right)}{(1 - x) + x (1 + \gamma)e^{s(x)}\nu_1/\nu_2}.
\label{eq:dsdx}
\eeq

\noindent The only way $s(x)$ can be non-monotonic is if $ds/dx = 0$ for some $x$ without $s(x)$ being constant.  Since the denominator of Eq.~\ref{eq:dsdx} is always positive, $ds/dx = 0$ only if $s(x) = \log(\nu_2/\nu_1)$ for some $x$.  However, if $s(x) = \log(\nu_2/\nu_1)$ for any $x$, then it must be constant at $\log(\nu_2/\nu_1)$ for all $x$.  We show this by substituting $s(x) = \log(\nu_2/\nu_1)$ into the saturation equation (Eq.~\ref{eq:saturation_condition_s}).  The $x$-dependence drops out and we are left with

\beq
\frac{\nu_2^{1/\gamma}}{\nu_1^{1 + 1/\gamma}} e^{\omega(1 + 1/\gamma)} = 1.
\label{eq:zero_dsdx}
\eeq

\noindent Therefore if the parameters satisfy this condition, then $s(x) = \log(\nu_2/\nu_1)$ for all $x$.  Therefore $ds/dx$ only equals zero when $s(x)$ is constant, and so $s(x)$ can never be a non-monotonic function of $x$.  

\begin{figure*}
\centering\includegraphics[width=\textwidth]{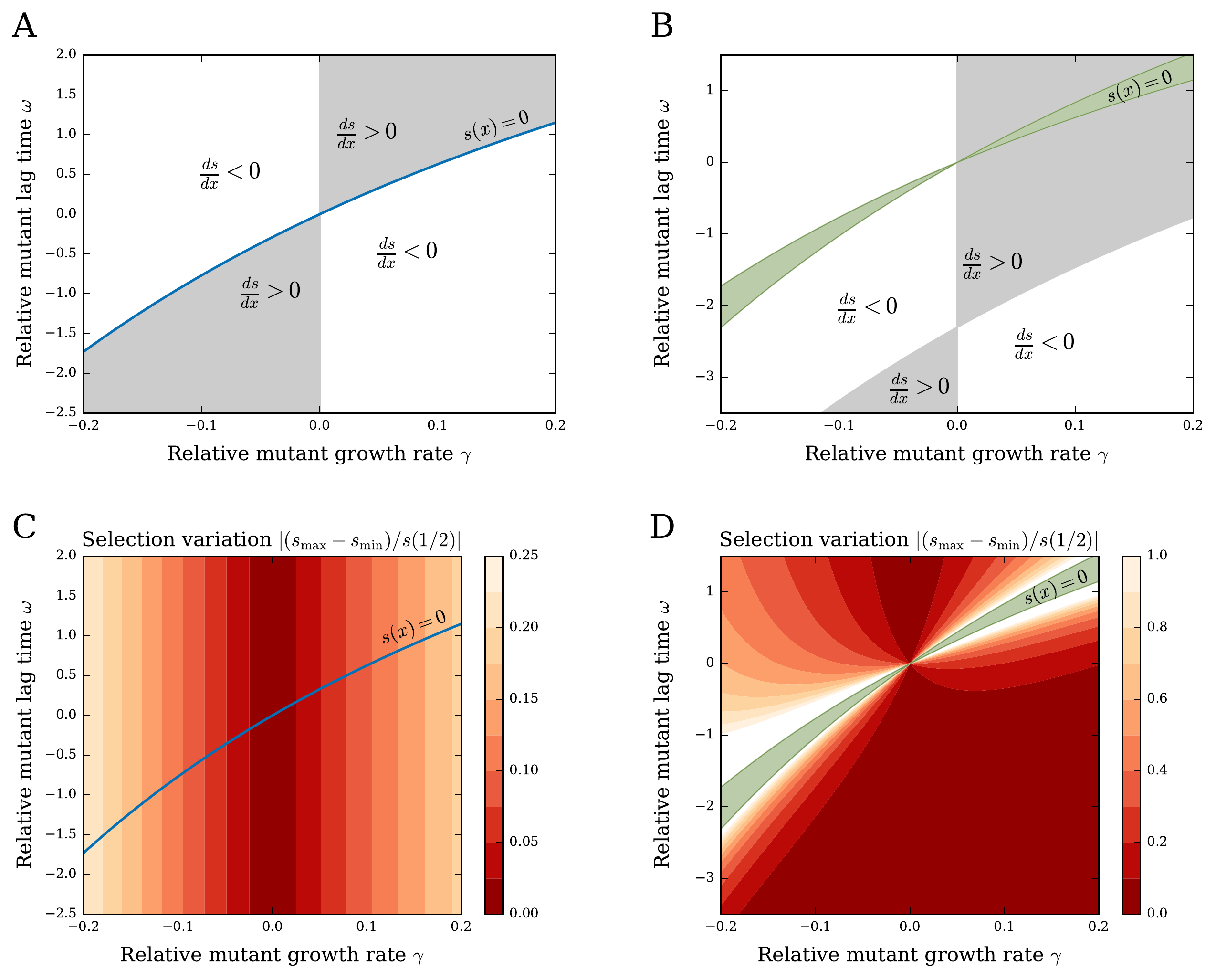}
\caption{
\textbf{Frequency-dependence of the selection coefficient over growth-lag trait space.}  
(A)~For a mutant and wild-type with equal yields ($\nu_1 = \nu_2 = 10^3$), the gray and white regions indicate where the selection coefficient $s(x)$ increases as a function of mutant frequency ($ds/dx > 0$) or decreases ($ds/dx < 0$).  The neutral boundary is in blue.  
(B)~Same as (A) but for a mutant and wild-type with unequal yields ($\nu_1 = 10^3$, $\nu_2 = 10^4$).  The conditionally-neutral region is shown in green.  
(C)~Relative variation of the selection coefficient over mutant frequencies when the mutant and wild-type have equal yields.  Yield values and the neutral boundary are the same as (A).  
(D)~Same as (C) but for a mutant and wild-type with unequal yields; yield values and the conditionally-neutral region are the same as (B).  The relative variation diverges in the conditionally-neutral region since $s(1/2) = 0$ for some points.
}
\label{fig:freq_dependence}
\end{figure*}

     Figure~\ref{fig:freq_dependence}A shows the sign of $ds/dx$ over growth-lag trait space for strains with equal yields ($\nu_1 = \nu_2$); Fig.~\ref{fig:freq_dependence}B shows the case of unequal yields ($\nu_1 \neq \nu_2$).  The boundaries between signs of $ds/dx$ are where $s(x)$ is a constant, and thus they are given by $\gamma = 0$ and Eq.~\ref{eq:zero_dsdx}.
Note that for equal yields, $s(x)$ is constant at zero along the neutral boundary (Fig.~\ref{fig:freq_dependence}A), whereas for unequal yields there is a separate boundary, away from the conditionally-neutral region, where $s(x)$ has a constant but nonzero value (Fig.~\ref{fig:freq_dependence}B).

     Another way to measure the frequency-dependence of selection is to consider its total variation across the whole range of frequencies.  We define the relative variation of selection as $|(s_{\max} - s_{\min})/s(1/2)|$, where $s_{\max}$ and $s_{\min}$ are the maximum and minimum values of $s(x)$ across all frequencies, and $s(1/2)$ is selection at the intermediate frequency $x = 1/2$.  
Since $s(x)$ is always a monotonic function of $x$, the maximum and minimum values are attained at the endpoints $x = 0$ and $x = 1$.  The selection coefficient is not technically defined for these values (since either the mutant or the wild-type is extinct), but we can determine its value in the limits $x \to 0$ and $x \to 1$.  In the limit of $x \to 0$, the saturation time must be the time for the wild-type alone to consume all the resources, and vice-versa for $x \to 1$:
     
\beq
\begin{split}
\lim_{x \to 0} t_\sat(x) & = \lambda_1 + \frac{1}{g_1}\log\left(\frac{RY_1}{N_0}\right), \\
\lim_{x \to 1} t_\sat(x) & = \lambda_2 + \frac{1}{g_2}\log\left(\frac{RY_2}{N_0}\right). \\
\end{split}
\eeq

\noindent Using the relationship between $s$ and $t_\sat$ in Eq.~\ref{eq:s_def_alternate2} and converting to dimensionless parameters (\eqparamdef), we have

\beq
\begin{split}
\lim_{x \to 0} s(x) & = \gamma\log \nu_1 - \omega(1 + \gamma), \\
\lim_{x \to 1} s(x) & = \left(\frac{\gamma}{1 + \gamma}\right)\log \nu_2 - \omega. \\
\end{split}
\eeq

\noindent Hence the total variation of selection coefficients is

\beq
\begin{split}
\left| s_{\max} - s_{\min} \right| & = \left| \lim_{x \to 1} s(x) - \lim_{x \to 0} s(x) \right| \\
& = \left|\gamma\left(\frac{\log\nu_2}{1 + \gamma} - \log\nu_1 + \omega \right)\right|. \\
\end{split}
\eeq

\noindent This result is exact (no weak selection approximation), but the approximate $s(x)$ expression in \eqsformula~gives an identical result.

     Normalizing this total range of selection by its magnitude at some intermediate frequency, such as $x = 1/2$, measures the relative variation in $s(x)$ over frequencies.  For equal yields ($\nu_1 = \nu_2$), the relative variation simplifies to

\beq
\left|\frac{s_{\max} - s_{\min}}{s(1/2)}\right| = \left| \frac{\gamma(2 + \gamma)}{2(1 + \gamma)} \right|.
\eeq
     
\noindent It is small over a large range of the trait space (Fig.~\ref{fig:freq_dependence}C), indicating that the frequency-dependence of selection is relatively weak for equal yields.  In contrast, when the yields are unequal ($\nu_1 \neq \nu_2$), the variation becomes very large near the conditionally-neutral region (Fig.~\ref{fig:freq_dependence}D).  This is because $s(1/2)$ goes to zero for some points in the conditionally-neutral region, while the total range $|s_{\max} - s_{\min}|$ remains finite.  Thus, the frequency-dependence of selection is most significant for mutants in the conditionally-neutral region; this is expected since these are the mutants that can coexist or be bistable with the wild-type.
     


\section{Robustness of coexistence to genetic drift}

     If the bottleneck population size $N_0$ at the beginning of each round is small, then stochastic effects of sampling from round to round (genetic drift) may be significant.  We can gauge the robustness of coexistence to these fluctuations by comparing the magnitude of those fluctuations, which is of order $1/N_0$, with $ds/dx$ measured at the coexistence frequency $\tilde{x}$ (Eq.~\ref{eq:dsdx}), which estimates the strength of selection for a small change in frequency around coexistence.  Coexistence will be robust against fluctuations if
     
\beq
\left| \frac{\gamma (1 - \nu_1/\nu_2)}{(1 - \tilde{x}) + \tilde{x}(1 + \gamma)\nu_1/\nu_2} \right| > \frac{1}{N_0}.
\eeq

\noindent This tells us the critical value of the bottleneck size $N_0$, which we can control experimentally, needed to achieve robust coexistence.  For example, if the mutant has 10\% slower growth rate ($\gamma = -0.1$) but 10\% higher yield ($\nu_2/\nu_1 = 1.1$), and coexistence occurs at $\tilde{x} = 1/2$, then $N_0$ must be greater than 100 for stabilizing selection at the coexistence frequency to be stronger than genetic drift.



\section{Fixation under frequency-dependent selection}

     If the population at the end of a competition round is randomly sampled to populate the next round, this is equivalent to a Wright-Fisher process with frequency-dependent selection coefficient $s(x)$ and effective population size $N_0$~\cite{Crow1970}.  
In the limit of a large population ($N_0 \gg 1$) and weak selection ($|s(x)| \ll 1$), the fixation probability of a mutant starting from frequency $x$ is

\beq
\phi(x) = \frac{\int_0^x dx'~ e^{2N_0 V(x')}}{\int_0^1 dx'~ e^{2N_0 V(x')}},
\label{eq:diffusion_phi}
\eeq

\noindent where $V(x)$ is the effective selection ``potential'':

\beq
V(x) = -\int_0^x dx'~ s(x').
\label{eq:V_def}
\eeq

\noindent This is defined in analogy with physical systems, where selection plays the role of a force and $V(x)$ is the corresponding potential energy function.  The mean time (number of competition rounds) to fixation, given that fixation eventually occurs, is

\begin{multline}
\theta(x) = \int_x^1 dx'~ \psi(x') \phi(x') (1 - \phi(x')) \\
+ \left(\frac{1 - \phi(x)}{\phi(x)}\right)\int_0^x dx'~ \psi(x') \left(\phi(x')\right)^2,
\label{eq:diffusion_tau}
\end{multline}

\noindent where

\beq
\psi(x) = \frac{2N_0e^{-2N_0 V(x)}}{x(1-x)} \int_0^1 dx'~ e^{2N_0 V(x')}.
\eeq

\noindent These results assume that mutations are rare enough to neglect interference from multiple \emph{de novo} mutations.  

     For simplicity we focus on the case of a single mutant cell (frequency $1/N_0$) at the beginning of a competition round.  To test the effect of frequency-dependence on fixation, we compare the true fixation probabilities and times, calculated from Eqs.~\ref{eq:diffusion_phi} and~\ref{eq:diffusion_tau} using $s(x)$ (\eqsformula), with the fixation probabilities and times predicted if selection has a constant value at $s(1/2)$, as is often measured in competition experiments~\cite{Elena2003}.  When selection is a constant across frequencies, Eq.~\ref{eq:diffusion_phi} simplifies to Kimura's formula~\cite{Crow1970}:
     
\beq
\phi(1/N_0) = \frac{1 - e^{-2s}}{1 - e^{-2N_0s}}.
\label{eq:Kimura_phi}
\eeq

\noindent Deviations from this relationship between $\phi$ and $s(1/2)$ are therefore indicative of significant frequency-dependence.

\begin{figure*}
\centering\includegraphics[width=\textwidth]{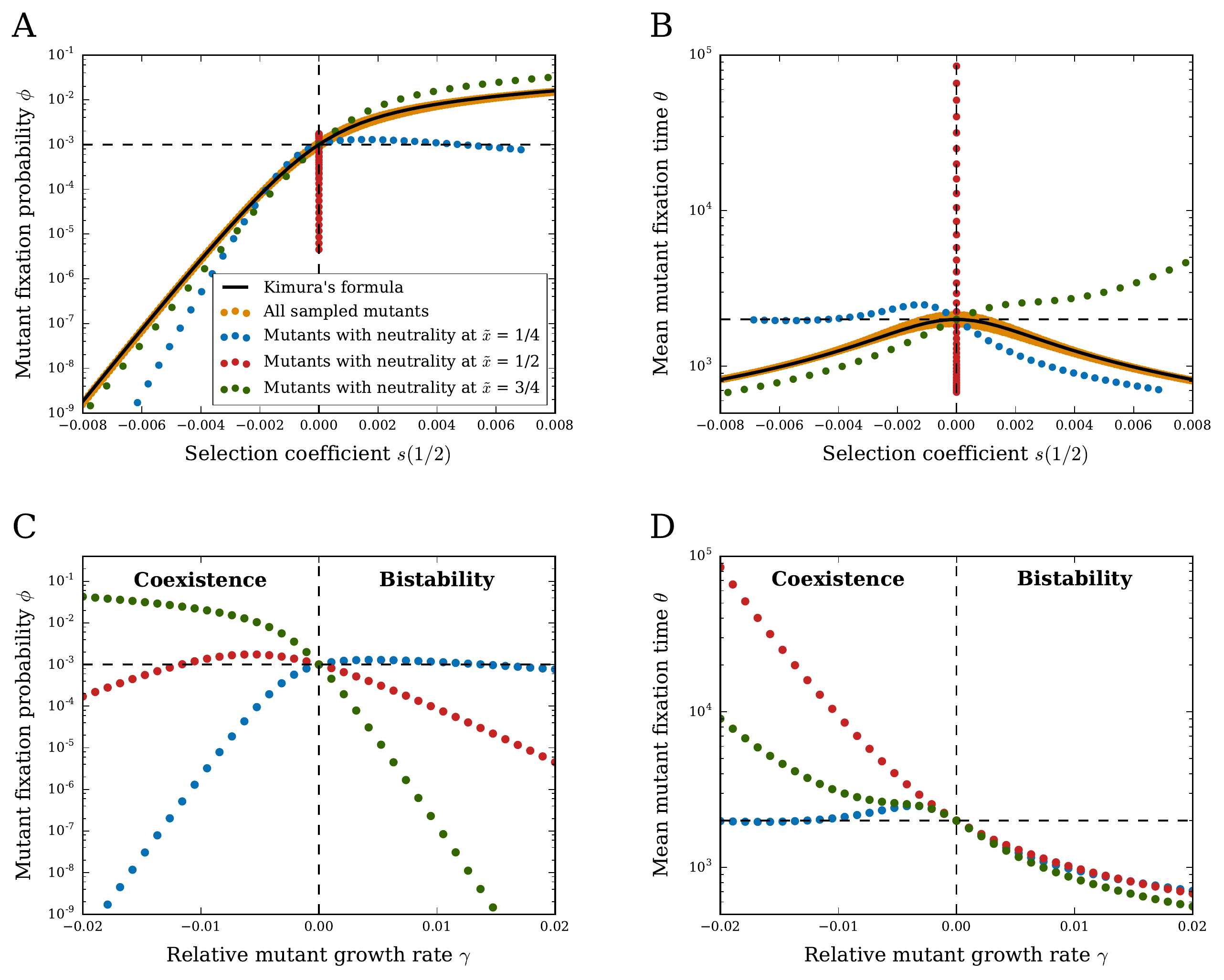}
\caption{
\textbf{Fixation probabilities and times of a mutant.}  
(A)~Fixation probability $\phi(1/N_0)$ as a function of the selection coefficient at frequency $x = 1/2$.  Orange points correspond to mutants uniformly sampled across a rectangular region of growth-lag trait space: $(\gamma, \omega) \in [-10^{-3}, 10^{-3}] \times [-5 \times 10^{-3}, 5 \times 10^{-3}]$.  Other points correspond to mutants with neutrality at specific frequencies (blue for $\tilde{x} = 1/4$, red for $\tilde{x} = 1/2$, green for $\tilde{x} = 3/4$).  We calculate fixation probabilities using the frequency-dependent selection coefficient $s(x)$ (\eqsformula) and Eq.~\ref{eq:diffusion_phi}; for comparison, the solid black line indicates the prediction from Kimura's formula (Eq.~\ref{eq:Kimura_phi}), assuming a frequency-independent selection coefficient.  The horizontal dashed line marks the neutral fixation probability $\phi = 1/N_0$.  
(B)~Same as (A), but with the mean fixation time $\theta(1/N_0)$ (conditioned on eventual fixation) on the vertical axis.  The solid black line marks the prediction for a frequency-independent selection coefficient (Eq.~\ref{eq:diffusion_tau}), and the horizontal dashed line marks the neutral fixation time $\theta = 2N_0$.  
(C)~Fixation probability $\phi(1/N_0)$ as a function of the relative growth rate $\gamma$ for conditionally-neutral mutants.  Colors indicate the same neutral frequencies as in (A) and (B).  Mutants with $\gamma < 0$ have coexistence, while mutants with $\gamma > 0$ are bistable (since $\nu_2 > \nu_1$).  Dashed lines are the same as in (A).  
(D)~Same as (C), but with the mean fixation time $\theta(1/N_0)$ on the vertical axis.  Dashed lines are the same as (B).   
In all panels, the relative yields are $\nu_1 = 10^3$ and $\nu_2 = 10^4$, and the initial population size is $N_0 = 10^3$.  
}
\label{fig:fixation}
\end{figure*}

     For several sets of mutants, Fig.~\ref{fig:fixation}A shows their measured selection coefficients $s(1/2)$ versus their fixation probabilities $\phi(1/N_0)$.  In orange are mutants obtained by uniformly scanning a rectangular region of growth-lag trait space (e.g., the trait space shown in Fig.~2A).  The black line shows the prediction from Kimura's formula (Eq.~\ref{eq:Kimura_phi}) assuming $s = s(1/2)$ is a constant selection coefficient for all frequencies; this frequency-independent approximation appears to describe these mutants well.  The mean fixation times $\theta(1/N_0)$ (Fig.~\ref{fig:fixation}B) for these mutants are also well-described by assuming constant selection coefficient $s(1/2)$.  This is because the frequency-dependence for these mutants is weak, as shown in Fig.~\ref{fig:freq_dependence}C,D.  Therefore a single measurement of the selection coefficient for these mutants at any initial frequency provides an accurate prediction of the long-term population dynamics.

     The plots of selection variation in Fig.~\ref{fig:freq_dependence}C,D indicate that the most significant frequency-dependence occurs for mutants in the conditionally-neutral region with unequal yields, i.e., mutants with coexistence \red{or bistability.}  We thus calculate the fixation probabilities and times for mutants with neutrality at particular frequencies, and compare these statistics to their selection coefficients at $x = 1/2$ as would be measured experimentally.  As expected, the fixation statistics show significant deviations from the predictions for constant selection.  In particular, mutants with neutrality at $\tilde{x} = 1/2$ (\eqcoexistence) have $s(1/2) = 0$ by definition, but they nevertheless show a wide range of fixation probabilities and times, some above the neutral values ($\phi = 1/N_0$, $\theta = 2N_0$) and some below.
     
     Figure~\ref{fig:fixation}C,D shows the fixation probabilities and times of conditionally-neutral mutants as functions of their relative growth rates $\gamma$, which separates mutants with coexistence from those with \red{bistability}:  the mutant has higher yield than that of the wild-type in this example ($\nu_2 > \nu_1$), so the mutants with worse growth rate ($\gamma < 0$) have coexistence while the mutants with better growth rate ($\gamma > 0$) are \red{bistable}.  
\red{Bistable mutants with a neutral frequency of} $\tilde{x} = 1/2$ fix with lower probability than would a purely neutral mutant (Fig.~\ref{fig:fixation}C), but if they do fix, they do so in less time (Fig.~\ref{fig:fixation}D).  We can understand this bistable case in analogy with diffusion across an energy barrier, using the effective selection potential defined in Eq.~\ref{eq:V_def}.  
The mutant starts at frequency $1/N_0$, and to reach fixation it must not only survive fluctuations from genetic drift while at low frequency, but it also must cross the effective selection potential barrier at the neutral frequency $\tilde{x}$.  Indeed, the mutant is actually deleterious at low frequencies (below the neutral frequency), and thus we expect the fixation probability to be lower than that of a purely neutral mutant.  If such a mutant does fix, though, it will do so rapidly, since it requires rapid fluctuations from genetic drift to cross the selection barrier.  
This effect is most pronounced for neutrality at relatively high frequencies; for low coexistence frequencies, such as $\tilde{x} = 1/4$, the barrier is sufficiently close to the initial frequency $1/N_0$ that it is easier to cross, and thus the fixation probability is closer to the neutral value (Fig.~\ref{fig:fixation}C).

     Mutants with coexistence, on the other hand, are described by a potential well at the neutral frequency. 
The fixation of these mutants is determined by a tradeoff between the initial boost of positive selection toward the neutral frequency, which helps to avoid immediate extinction, and the stabilizing selection they experience once at coexistence.  
In particular, once at the neutral frequency, the mutant must eventually cross a selection barrier to reach either extinction or fixation. 
However, the barrier to fixation is always higher, and thus the mutant has a greater chance of going extinct rather than fixing.  As we see for mutants with coexistence at $\tilde{x} = 1/2$, decreasing $\gamma$ from zero initially improves the probability of fixation over neutrality, but eventually it begins to decrease.  Thus, the frequency-dependence of mutants with coexistence plays a crucial role in shaping their fixation statistics, and their ultimate fates depend crucially on their individual trait values (i.e., $\gamma$).

\section{Additivity of the selection coefficient}

     The additivity condition (\eqadditivity) is approximately satisfied if strains $i$, $j$, and $k$ have only small differences in growth rates, lag times, and yields.  Conceptually, this is because the saturation times $t_\sat$ for each binary competition between pairs of strains are all approximately equal, but we can also show this directly using the selection coefficient formula.  Let $\gamma_{ij} = (g_i - g_j)/g_j$, $\omega_{ij}= (\lambda_i - \lambda_j)g_j$, and $\mu_{ij} = (\nu_i - \nu_j)/\nu_j$ be the relative differences in growth rate, lag time, and yield for strains $i$ and $j$.  If these relative differences are all small, then they each approximately obey the additivity condition across strains:
     

\begin{align}
\label{eq:trait_change_additivity}
\gamma_{ik} & = (1 + \gamma_{ij})(1 + \gamma_{jk}) - 1 && \approx \gamma_{ij} + \gamma_{jk}, \nonumber\\
\omega_{ik} & = \frac{\omega_{ij}}{1 + \gamma_{jk}} + \omega_{jk} && \approx \omega_{ij} + \omega_{jk}, \\
\mu_{ik} & = (1 + \mu_{ij})(1 + \mu_{jk}) - 1 && \approx \mu_{ij} + \mu_{jk}. \nonumber
\end{align}
     
\noindent In this same limit, the total selection coefficient for strains $i$ and $j$ is approximately
     
\beq
s_{ij} \approx \gamma_{ij} \log \nu_j - \omega_{ij}.
\eeq

\noindent Note that, to leading order, the change in yield $\mu_{ij}$ does not appear.  Using Eq.~\ref{eq:trait_change_additivity} and $\nu_j = (1 + \mu_{jk})\nu_k$, we have 

\beq
\begin{split}
s_{ij} + s_{jk} & \approx \gamma_{ij} \log \nu_j - \omega_{ij} + \gamma_{jk} \log \nu_k - \omega_{jk} \\
& \approx \gamma_{ik} \log \nu_k - \omega_{ik} \\
& \approx s_{ik}. \\
\end{split}
\eeq

\noindent Therefore the selection coefficient is approximately additive when differences between traits are small.



\section{Transitivity of the selection coefficient}

     Since we are only concerned with the sign of selection in determining transitivity, we focus on the signed component of the selection coefficient in \eqsformula.  It is also more convenient to use growth times $\tau_i = 1/g_i$ rather than growth rates, and the quantity $h_{ij} = \log \frac{1}{2}H(\frac{\nu_j}{1-x}, \frac{\nu_i}{x})$ for the logarithm of the harmonic mean yield.  We define the signed component of the selection coefficient for strain $i$ over strain $j$ to be

\beq
(\tau_j - \tau_i)h_{ij} + \lambda_j - \lambda_i.
\eeq

\noindent That is, $s_{ij}$ is proportional to this quantity up to an overall factor that is always nonnegative.

     We first consider whether neutrality is a transitive property of strains.  Three strains are all pairwise neutral if their traits satisfy
     
\beq
\begin{split}
(\tau_1 - \tau_2)h_{12} + \lambda_1 - \lambda_2 & = 0, \\
(\tau_2 - \tau_3)h_{23} + \lambda_2 - \lambda_3 & = 0, \\
(\tau_3 - \tau_1)h_{13} + \lambda_3 - \lambda_1 & = 0. \\
\end{split}
\label{eq:transitive_neutrality}
\eeq

\noindent If all three strains have equal yields $\nu_1 = \nu_2 = \nu_3$ ($h_{12} = h_{23} = h_{13}$ for all frequencies), then any two of these equations imply the third (e.g., by adding them together), which means that neutrality is transitive when all strains have equal yields.  If two of the yields are equal while the third is distinct, then transitivity only holds if two of the strains are identical (equal growth and lag times).  For example, if $\nu_1 = \nu_2 \neq \nu_3$, then we can add together the last two equations in Eq.~\ref{eq:transitive_neutrality} to obtain

\beq
(\tau_2 - \tau_1)h_{13} + \lambda_2 - \lambda_1 = 0,
\eeq

\noindent (using $h_{23} = h_{13}$), but this is only consistent with the first equation in Eq.~\ref{eq:transitive_neutrality} if $\tau_1 = \tau_2$ and $\lambda_1 = \lambda_2$, i.e., strains 1 and 2 are identical in all traits.

     If all the yields have distinct values, then transitivity will generally not hold for arbitrary values of the growth traits.  However, it is still \emph{possible} for three strains with distinct yields to all be pairwise neutral, but only with very specific values of the traits.  Note that with unequal yields, neutrality at all frequencies is not possible, so pairs of strains are only conditionally-neutral, with neutrality at some particular frequency.  These frequencies are encoded in the quantities $h_{ij}$.  We thus fix the yields and the desired neutral frequencies to arbitrary values, and without loss of generality, we can assume $h_{12} < h_{13} < h_{23}$ (e.g., by putting the strains in order of increasing yields).  We can also choose any values of $\tau_1$ and $\lambda_1$ since this amounts to a rescaling and shift of time units.  Therefore we are left with three linear equations (Eq.~\ref{eq:transitive_neutrality}) for four unknowns: $\tau_2, \tau_3, \lambda_2, \lambda_3$.  If we choose any value of the strain 2 growth time that obeys 
     
\beq
\tau_2 > \left(\frac{h_{13} - h_{12}}{h_{23} - h_{12}}\right) \tau_1
\label{eq:tau2_min}
\eeq

\noindent (note the factor in parentheses is always positive by assumption), then Eq.~\ref{eq:transitive_neutrality} has a unique solution for the remaining quantities:

\beq
\begin{split}
\tau_3 & = \frac{\tau_2(h_{23} - h_{12}) - \tau_1(h_{13} - h_{12})}{h_{23} - h_{13}}, \\
\lambda_2 & = (\tau_1 - \tau_2)h_{12} + \lambda_1, \\
\lambda_3 & = (\tau_1 - \tau_2)\left(\frac{h_{23} - h_{12}}{h_{23} - h_{13}}\right)h_{13} + \lambda_1. \\
\end{split}
\label{eq:tau3_lambda2_lambda_3_solutions}
\eeq

\noindent The linear system actually has a unique solution regardless of Eq.~\ref{eq:tau2_min}, but without that condition $\tau_3$ may be negative.  Therefore a set of three strains with unequal yields can all be pairwise conditionally-neutral only if the growth traits for strains 2 and 3 satisfy Eqs.~\ref{eq:tau2_min} and~\ref{eq:tau3_lambda2_lambda_3_solutions}.  For example, in this manner one can construct three strains that all coexist in pairs.

     We now turn to constructing sets of three strains such that there is a nontransitive cycle of selective advantage in binary competitions, i.e., strain 2 beats strain 1 in a binary competition, strain 3 beats strain 2, but strain 1 beats strain 3.  Therefore the growth traits of the three strains must satisfy
     
\beq
\begin{split}
(\tau_1 - \tau_2)h_{12} + \lambda_1 - \lambda_2 & > 0, \\
(\tau_2 - \tau_3)h_{23} + \lambda_2 - \lambda_3 & > 0, \\
(\tau_3 - \tau_1)h_{13} + \lambda_3 - \lambda_1 & > 0. \\
\end{split}
\label{eq:nontransitive_selection}
\eeq

\noindent All three yields cannot be equal; if they are, adding together any two of the inequalities in Eq.~\ref{eq:nontransitive_selection} gives an inequality that is inconsistent with the third one.  Otherwise, the three yields can take arbitrary values, including two of them being equal.  Since we can cyclically permute the strain labels, without loss of generality we assume strain 1 has the smallest yield ($\nu_1 < \nu_2, \nu_3$).  Therefore the harmonic mean logarithms obey $h_{23} > h_{12}, h_{13}$.  We can also choose any values of $\tau_1$ and $\lambda_1$ as before.

     We must now choose the growth traits of strains 2 and 3 ($\tau_2, \tau_3, \lambda_2, \lambda_3$) to satisfy the inequalities of Eq.~\ref{eq:nontransitive_selection}.  We use a geometrical approach to understand the available region of trait space for these strains.  The lag time for strain 3 is bounded from above and below according to (combining the second and third inequalities in Eq.~\ref{eq:nontransitive_selection})

\beq
(\tau_1 - \tau_3)h_{13} + \lambda_1 < \lambda_3 < (\tau_2 - \tau_3)h_{23} + \lambda_2.
\label{eq:lambda3_condition}
\eeq

\noindent The upper and lower bounds are both functions of $\tau_3$.  The upper bound will be above the lower bound as long as $\tau_3$ satisfies

\beq
\tau_3 < \frac{\tau_2 h_{23} - \tau_1 h_{13} + \lambda_2 - \lambda_1}{h_{23} - h_{13}}.
\label{eq:tau3_condition}
\eeq

\noindent Since $\tau_3$ must be positive, this upper bound of $\tau_3$ must also be positive.  The denominator of the right-hand side of Eq.~\ref{eq:tau3_condition} is positive by assumptions about the yields, so therefore the numerator must be positive as well.  This leads to a lower bound on the lag time of strain 2 $\lambda_2$; we can combine this with an upper bound on $\lambda_2$ from the first equation of Eq.~\ref{eq:nontransitive_selection} (strain 2 beats strain 1) to obtain 

\beq
\tau_1 h_{13} - \tau_2 h_{23} + \lambda_1 < \lambda_2 < (\tau_1 - \tau_2)h_{12} + \lambda_1.
\label{eq:lambda2_condition}
\eeq

\noindent Finally, the upper bound for $\lambda_2$ will be above the lower bound as long as $\tau_2$ satisfies

\beq
\tau_2 > \max\left(\left(\frac{h_{13} - h_{12}}{h_{23} - h_{12}}\right)\tau_1, 0\right).
\label{eq:tau2_condition}
\eeq

\noindent Altogether, we can construct a set of nontransitive strains by choosing any yields $\nu_1, \nu_2, \nu_3$ satisfying $\nu_1 < \nu_2, \nu_3$, and any values for the growth traits $\tau_1, \lambda_1$ of strain 1; we then choose $\tau_2$ according to Eq.~\ref{eq:tau2_condition} and $\lambda_2$ according to Eq.~\ref{eq:lambda2_condition}; finally, we choose $\tau_3$ according to Eq.~\ref{eq:tau3_condition} and $\lambda_3$ according to Eq.~\ref{eq:lambda3_condition}.

\bibliographystyle{unsrt}
\bibliography{References}